\documentclass{aa}
\usepackage{graphicx}
\usepackage{longtable}
\usepackage{latexsym}
\usepackage{amssymb}
\usepackage{lscape}
\input{psfig.txt}
\begin{document}
   \title{An extragalactic HII region in the Virgo cluster\thanks{Based on observations obtained with the Loiano 
   telescope belonging to the University of Bologna (Italy) 
   and with the Calar Alto observatory operated by the Centro Astronomico
	Hispano Aleman (Spain).}}
   %\subtitle{}

   \author{L.Cortese\inst{1}, G.Gavazzi\inst{1}, A.Boselli\inst{2} \and  J.Iglesias-Paramo\inst{2}}

   \offprints{L.Cortese}

   \institute{Universit\'{a} degli Studi di Milano-Bicocca, P.zza della Scienza 3, 20126 Milano, Italy.\\
              \email{Luca.Cortese@mib.infn.it; Giuseppe.Gavazzi@mib.infn.it}
         \and
             Laboratoire d'Astrophysique de Marseille, BP8, Traverse du Siphon, F-13376 Marseille, France.\\
             \email{jorge.iglesias@oamp.fr; alessandro.boselli@oamp.fr}
             }

   \date{Received 21 July 2003 | Accepted 2 December 2003}

\abstract{We present spectroscopic observations for six  emission-line objects projected onto the Virgo cluster. 
These sources have been selected from narrow band ($\rm H\alpha+[NII]$) images showing faint 
detectable continuum emission and $EW>100\rm \AA$.
Five of these sources result $\rm [OIII] \lambda 5007$ emitters at $z \sim 0.31$, while one 
\emph{122603+130724} is confirmed to be an HII region belonging to the Virgo cluster. 
This point-like source has a recessional velocity of $\rm \sim 200~km~s^{-1}$, 
and is associated with the giant galaxy VCC873 (NGC 4402).
It has a higher luminosity, star formation rate and metallicity than the extragalactic HII region recently 
discovered near the Virgo galaxy VCC836 by
Gerhard et al. (\cite{gerhard}). 

}

\titlerunning{An extragalactic HII region in the Virgo cluster}
\authorrunning{L.Cortese et al.}

   \maketitle
%
%________________________________________________________________

\section{Introduction}

The presence of a diffuse intra-cluster light component was for the first time proposed by 
Zwicky(\cite{zwicky}), who detected a light excess between the galaxies in the 
Coma cluster.
In the last twenty years, the presence of intra-cluster starlight (ICL) has been inferred 
from surface brightness measurements 
(Bernstein et al. \cite{bernstein}) and from the detection of individual extragalactic stars 
(Ferguson et al. \cite{ferguson}), planetary nebulae (Arnaboldi et al. \cite{arnaboldi96}), 
and HII regions (Gerhard et al. \cite{gerhard}). 
The study of ICL and its origin could shed light on the evolution of galaxies in clusters,
a complex process involving various mechanism such as tidal interactions (Gnedin \cite{gnedin}), 
galaxy harassment (Moore et al. \cite{moore}) and ram pressure (Gunn \& Gott \cite{gunn}). 
Gravitational interactions between cluster galaxies may strip gas and stars from these systems contributing to  
the "pollution" of the intra-cluster medium (ICM) and forming, in some cases, extragalactic HII regions as the 
ones recently observed in the Abell cluster 1367 
(Sakai et al. \cite{sakai}, Gavazzi et al. \cite{gavazzi03}).
However the interstellar medium may also  be removed from fast-moving galaxies by the ram pressure mechanism
(Quilis et al. \cite{quilis}, Gavazzi et al. \cite{gavazzi01}). 
Some of this gas may collapse and form stars, which would contribute to the ICL. 
Moreover extragalactic massive stars might explode as supernovae enriching the ICM.
Thus the ICL properties should provide important clues not only on the dynamical state 
of clusters and galaxies (Miller \cite{miller}), but also on their chemical evolution.
However, only one intra-cluster HII region candidate has been confirmed spectroscopically 
so far (Gerhard et al. \cite{gerhard}).\\   
In this work we present the result of spectroscopic observations of a second extragalactic 
HII region in the Virgo cluster, confirming its discovery by Cortese et al. (\cite{cortese}).\\
We assume for the Virgo cluster a distance modulus of 31.15, corresponding to a distance of 17 Mpc.

\section{Sample selection}
With the aim of detecting faint emission line objects in the Virgo cluster region, we visually 
inspected the net ($\rm H\alpha+[NII]$) frames obtained at the Isaac Newton Telescope by Boselli et al. (\cite{boselli}).
They observed nine different regions centered on BCD galaxies in the Virgo cluster, covering a total area 
of $\sim$2.5 square degrees with an $\rm H\alpha$ flux limit $\rm \sim10^{-15} ~erg~s^{-1}~cm^{-2}$. 
Since the seeing conditions during the observations varied field-to-field and even between ON and OFF-band  
observations, a conservative criterion was assumed for discriminating real from spurious detections, namely,   
we consider real those sources with $EW>100~\rm \AA$ and with a $S/N$ ratio $\rm >$ 5 in the net frame. 
Only six objects were found to meet this criterion.
As shown in Table \ref{Tab3}, they are all faint sources with detectable 
$r'$ continuum emission ($20.1 \leq m_{r'}\leq 21.6$) suggesting
that they are not extragalactic planetary nebulae, since these types of sources have no continuum emission 
(Arnaboldi et al. \cite{arnaboldi03}, Feldmeier et al. \cite{feldmeier}).
The emitting line systems detected in the narrow band filter could be extragalactic HII regions at the distance of 
the Virgo cluster, as the one recently discovered by Gerhard et al. (\cite{gerhard}) or background emission-line galaxies. 
With the aim of understanding the real nature of these sources we undertook a spectroscopic follow-up.

\section{Observations and data reduction}
Long-slit, low dispersion spectra were obtained in several observing runs since 2002 using the imaging spectrograph 
BFOSC attached to the Cassini 1.5m telescope at Loiano (Italy), and  with CAFOS attached to the 2.2m 
telescope of the Calar Alto Observatory (Spain). 
Table \ref{Tab1} lists the characteristics of the instrumentation in the adopted set-up.
The journal of observations is presented in Table \ref{Tab2}.\\
The observations at Loiano were performed using a 1.5 or 2.5 arcsec slit, depending on the seeing conditions, and the 
wavelength calibration was secured with exposures of HeAr lamps. 
The observations at Calar Alto were carried out with a 1.5 arcsec slit and the wavelength calibration was carried out 
with exposures of CdHe lamps. 
In all runs the observations were obtained in nearly photometric conditions or with thin cirrus.\\ 
The data reduction was performed in the IRAF\footnote{Image Reduction and Analysis Facility is written and supported by the National 
Optical Astronomy Observatories (NOAO) in Tucson, Arizona. 
NOAO is operated by the  Association of Universities for Research in Astronomy (AURA), 
Inc. under cooperative agreement with the National Science Foundation.} 
environment. After bias subtraction, 
the cosmic rays were removed using the task COSMICRAYS and/or by visual inspection. 
The lamp wavelength calibration was checked against known sky lines. 
These were found within $\rm \sim 1~\AA$ from their nominal positions, providing an estimate of the systematic 
uncertainty on the derived velocities of $\rm \sim 50~km~s^{-1}$.
Sky subtraction was performed on the 2D spectra using the IRAF/LONGSLIT task BACKGROUND. 
BACKGROUND was used to fit the 2D spectra row-by-row with a first order Legendre function with $3 \sigma$  
rejection iteration. The sky was fitted in two windows of $\rm \sim 100~ pixels$, one on each side of the detected emission lines.
After subtraction of sky background, one-dimensional spectra were extracted from the frames using $APSUM$. 
The apertures were limited to regions where the signal intensity was above 1 $\sigma$ of the sky noise.
For the source \emph{122603+130724} observations of the star FEIGE34 provided the flux calibration. 
The spectrophotometric standard was observed in nearly photometric conditions.

\section{Results}

\begin{figure*}[!b]
\centering
\includegraphics[width=15.1cm]{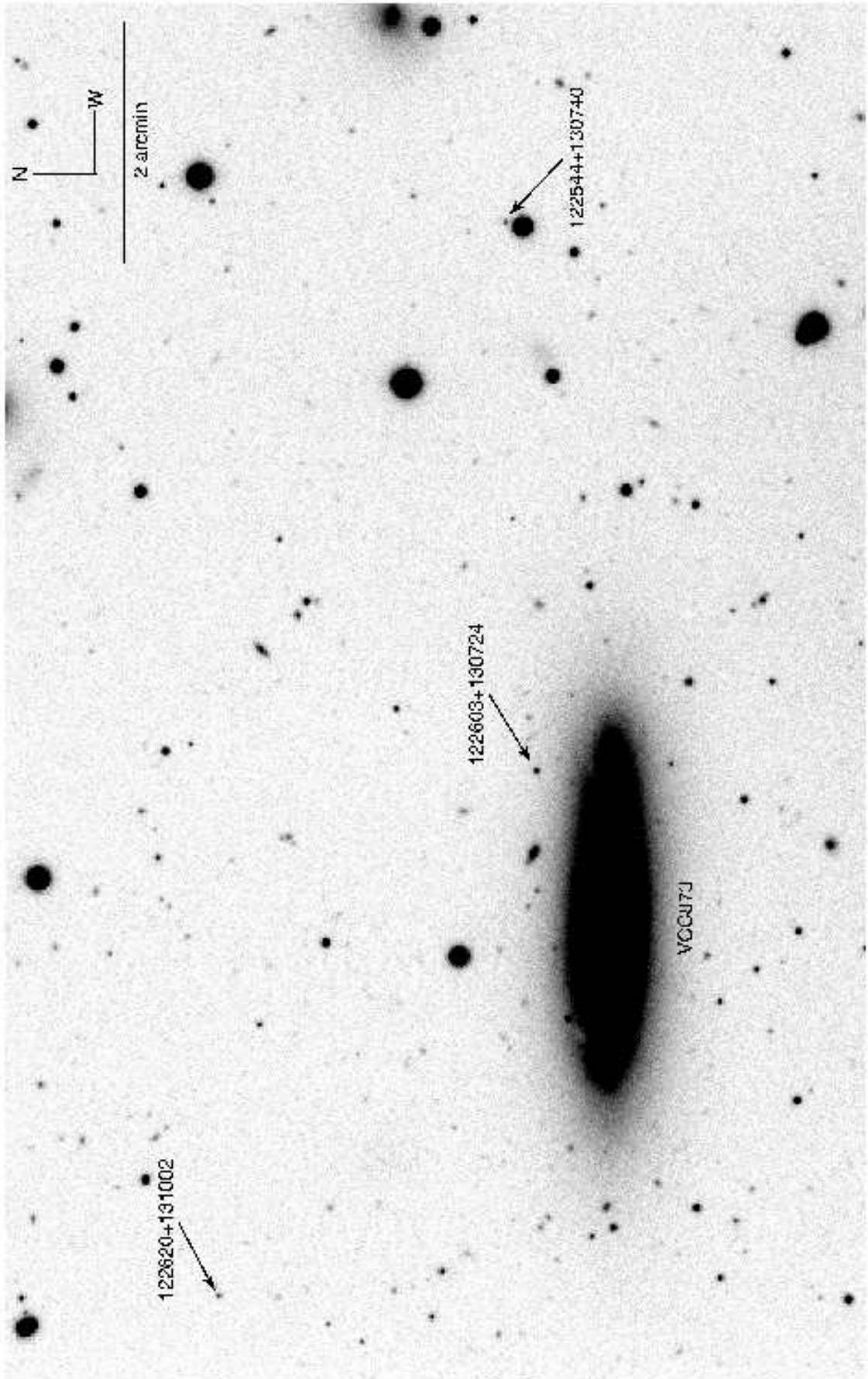}
\caption{Three of the emission-line objects  (\emph{122603+130724}, \emph{122544+130740}, \emph{122620+131002}) 
appear clustered around VCC873, as seen in the ON-band $\rm H_\alpha$ image.}
\label{vcc873}
\end{figure*}

\begin{figure*}
%\centering
\includegraphics[width=19cm]{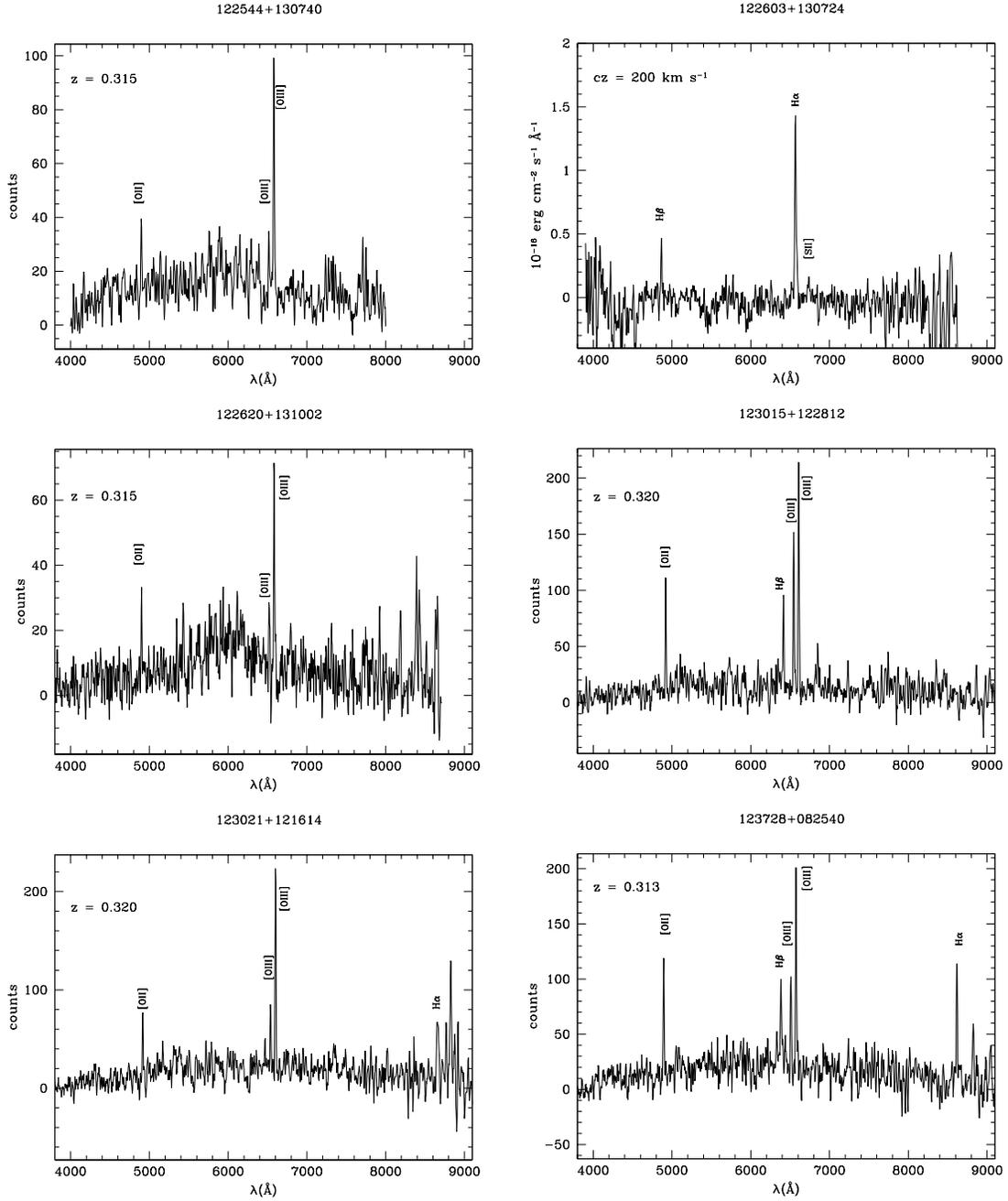}
\caption{The observed spectra. The galaxy identification, the emission line detected and the recessional velocity are labeled on each panel. 
Only the spectrum of \emph{122603+130724} is flux calibrated.}
\label{spectra}
\end{figure*}

The results of the measurements obtained in this work are listed in Table \ref{Tab3}. 
%as follow:\\
%Column 1: Object name.\\
%Column 2, 3: (J2000) celestial coordinates, measured with a few arcsec uncertainty.\\
%Column 4: $r'$ band magnitude.\\
%Column 5: Observed recessional velocity.\\
%Column 6: Emission line detected in the spectrum.\\
The obtained spectra are shown in Fig \ref{spectra}.
Five sources were found to be $\rm [OIII]\lambda5007$ emitters at $z \sim 0.31$, 
only one source, \emph{122603+130724}, belongs to the Virgo cluster.
Cortese et al. (\cite{cortese}) in their preliminary analysis, correctly interpreted
the single emission line detected in the spectrum of \emph{122603+130724} as due to $\rm H\alpha$ at the redshift of
$\rm 200~km~s^{-1}$, consistent with the velocity of VCC873 ($\rm 232~km~s^{-1}$).
Conversely the tentative interpretation that \emph{122544+130740} might be 
another extragalactic HII region is ruled out by the present observations which show that the source is in fact
a background galaxy and that the single emission line detected by Cortese et al. (\cite{cortese}) is $\rm [OIII]\lambda5007$  at $z = 0.315.$\\
\emph{122603+130724} lies at a projected distance of 3 kpc from  VCC873 (see Fig.\ref{vcc873}).
Boselli et al. (\cite{boselli}) observed this bright galaxy in the narrow-band $\rm H\alpha$ filter and in the broad-band 
$r'$ and B filters. Spectrophotometric standards were observed only for  $\rm H\alpha$ and B calibration.
However from the narrow-band calibration we were able to calibrate the $r'$ frame too.
The emitting line object \emph{122603+130724} is clearly detected in both broad-band images as 
a point-like source with $m_{B} \sim 22.25$ and $m_{r'} \sim 21.60$.
The uncertainties associated with these measures are $\sim 0.25$ mag due to contamination by VCC873.
Not surprisingly the continuum is not detected in the spectrum of Fig. \ref{spectra}
because $m_{r'} \sim 21.60$ corresponds to a flux of $f_{r'}\sim 6~10^{-18}~\rm erg~s^{-1}~cm^{-2}~\AA^{-1}$,
below the sensitivity limit $f_{r'}\sim 10^{-17}~\rm erg~s^{-1}~cm^{-2}~\AA^{-1}$ of our observations.\\
From the narrow-band images we measured $\rm H\alpha+[NII]$ flux $\rm = 10^{-14.56 \pm 0.05}~erg~s^{-1}~cm^{-2}$ and 
EW($\rm H\alpha+[NII]$) $\rm \sim 700 \pm 80 \AA$.
The uncertainty on the $\rm H\alpha+[NII]$ flux includes three contributions: 
the Poisson photon counts error, 
the uncertainty on the background and the photometric uncertainty, which
is assumed as 10\% of the net flux. These errors were determined separately on
the ON and OFF-band frames, and combined using the standard error propagation.   
The error in the equivalent width, is computed similarly to
the flux uncertainty, except that the error on the absolute flux scale
does not affect the equivalent width.\\
The flux-calibrated spectrum clearly shows three emission lines at the redshift of Virgo: 
$\rm H\alpha$, $\rm H\beta$ and $\rm [SII]\lambda6717,6731$.
The $\rm H\alpha$ line is blended with $\rm [NII]\lambda6583$. 
Using the task SPLOT we deblended the two components providing a crude estimate of the line ratio 
$\rm[NII]\lambda6583/H\alpha\sim 0.17$. 
The $\rm H\alpha$ flux obtained from the narrow-band images is 
consistent with the flux $f_{H\alpha} \sim 10^{-14.55}~erg~s^{-1}~cm^{-2}$ measured from the spectrum.
Assuming a theoretical value of 2.86 for the Balmer decrement ($T=10^4$ K and $n_e = 100~ \rm e/cm^{3}$), 
we obtain $A(H\beta) = 1.67$ and correct the line fluxes using the dereddening 
law of Cardelli et al. (\cite{cardelli}).
The dereddened $\rm H\alpha$ flux is $f_{H\alpha} \sim 10^{-14.10}~erg~s^{-1}~cm^{-2}$, 
translating into a luminosity $L_{H\alpha} \sim2.87~ 10^{38} \rm ~erg~s^{-1}$.
%We evaluated the Balmer decrement from the ratio $\rm H\alpha/H\beta$ assuming the theoretical value of 2.86 
%for $T=10^4$ K and $n_e = 100~ \rm e/cm^{3}$. 
%We obtained $A(H\beta) = 1.67$, and corrected the line fluxes using the dereddening law of Cardelli et al. (\cite{cardelli}).
The $\rm [OII]\lambda3727, [OIII]\lambda5007~ and ~4959$ lines are undetected but we derive an upper limit to 
their intensity.
%We did not detect the $\rm [OII]\lambda3727, [OIII]\lambda5007~ and ~4959$ lines, but we derived an upper limit to 
%their intensity.
The observed and reddening corrected line fluxes are listed in Table \ref{lines}.
We use the detected emission lines $\rm [NII]\lambda6583$ and $\rm[SII]\lambda6717,6731$  
to obtain an estimate of the metallicity  
of \emph{122603+130724} (Kewley \& Dopita \cite{kewley}).
Unfortunately the ratio $\rm [NII]\lambda6583/[SII]\lambda6717,6731$  
is strongly dependent on the ionization parameter (see Fig. 4 in Kewley \& Dopita \cite{kewley}), 
thus we can only derive a lower limit $\rm 8.6\leq 12 + log(O/H)$, to the abundance of \emph{122603+130724}.\\
The total number of H-ionizing photons estimated from (Osterbrock \cite{osterbrock}) is:
\begin{displaymath}
Q(H^0) = \frac{\alpha_B}{\alpha_{H\alpha}^{eff}} \times \frac{L_{H\alpha}}{h\nu_{H\alpha}} \sim 2.8 \times~10^{50}~\rm s^{-1} 
\end{displaymath}
where $\rm \alpha_B/\alpha_{H\alpha}^{eff} \sim 2.96$.\\
Thus the total mass and size of the gas cloud are:
\begin{displaymath}
M_{HII} = Q(H^0)\frac{m_p}{n_e \alpha_B} \sim 5150~ (\frac{n_e}{100~ \rm cm^{-3}})^{-1}  \rm~M_{\odot} 
\end{displaymath}
and
\begin{displaymath}
r_{HII} = \Big(\frac{M_{HII}(1+y^+)}{(4\pi/3)n_e m_p}\Big)^{1/3} \sim 8.2~ (\frac{n_e}{100~ \rm cm^{-3}})^{-2/3} \rm~pc      
\end{displaymath}
where $n_e$ is the electron density, $m_p$ the proton mass and $y^+ \sim 0.1$ is the fraction of ionized helium.\\
The total number of H-ionizing photons and the mass of the gas cloud are typical of giant HII regions (Stasinska \cite{stasinska}). 
Moreover the detectable continuum emission of \emph{122603+130724} and the low value of the ratio 
$\rm [OIII]\lambda5007 / H\alpha \leq 0.6$ rule out the possibility that this source is an intra-cluster planetary nebulae, 
confirming the hypothesis that it is an extragalactic HII region associated with the bright galaxy VCC873.\\
Recently Gerhard et al. (\cite{gerhard}) found 17 candidate extragalactic HII regions in the Virgo cluster, among which 
only one, associated with VCC836, was confirmed spectroscopically. 
This source differs from the HII region presented in this work as it is smaller ($r_{HII} \sim 3.5~ \rm pc$), 
has a lower luminosity ($L_{H\alpha} \sim1.3~ 10^{37} \rm ~erg~s^{-1}$) and metallicity ($\rm 12+ log(O/H) \sim 8.1$) than \emph{122603+130724}. 
Moreover the two giant galaxies to which the HII regions are associated have different properties.
They are both edge-on star forming spiral galaxies but, while VCC836 
is a Seyfert2 galaxy with a very extended emission-line region (Yoshida et al. \cite{yoshida}), 
VCC873 is a non active object with no evidence of extended extra-planar $\rm H\alpha$ emission.\\ 
This evidence indicates the possibility that the two extragalactic HII regions could have different origins. 
Gerhard et al. (\cite{gerhard}) suggest that 
the HII region associated with VCC836 might be a tidal debris of a past interaction with a gas-rich galaxy, as the very extended emission-line 
region associated with this bright galaxy indicates (Yoshida et al. \cite{yoshida}).
However Vollmer \& Huchtmeier (\cite{vollmer}) and Yoshida et al. (\cite{yoshida03}) propose
an alternative interpretation in which the extended emission-line regions, and perhaps the extragalactic HII 
region, are formed by the interaction of VCC836 with the dense ICM of the Virgo cluster.\\
The mechanism responsible for the formation of the extragalactic HII region \emph{122603+130724} 
is still not well understood, however the absence of a clear signature of interaction 
of VCC873 with another system rules out a tidal stripping scenario.\\ 
In recent years, deep HI observations of edge-on nearby spiral galaxies outlined 
the presence of a thick component (halo) of neutral hydrogen, with rotational velocity lower than the disk component, 
up to distances of 10-15 kpc from the disk (Swaters et al. \cite{swaters}).
Moreover large-scale radial inflow toward the galaxy center was revealed (Fraternali et al. \cite{fraternali}).
These observations suggest a complex gas circulation between the disk and the halo of star-forming spiral 
galaxies, supporting the galactic fountain model (e.g. Shapiro \& Field \cite{shapiro}; Bregman \cite{bregman}).
This model predicts that the gas, ionized by supernova explosion and stellar winds, leaves the plane of the disk.
After cooling, the ejected gas, presumably concentrated in clouds, falls back
into the disk.\\
\emph{122603+130724} lies within the halo of VCC873.  
Its velocity differs only by $\rm \sim50~km~s^{-1}$ from the velocity observed in the disk of VCC873 
near its location (Rubin et al. \cite{rubin}), suggesting that \emph{122603+130724} is 
gravitationally bound to the bright galaxy. 
The line ratio $\rm [NII]\lambda6583/[SII]\lambda6717,6731 \sim 1.2$ of \emph{122603+130724} is consistent 
with the metallicity obtained from the drift-scan mode spectrum of VCC873 itself, published by Gavazzi et al. (\cite{goldmine}, \cite{gavazzi04}), 
suggesting that this HII region formed from enriched material ejected from VCC873.\\
The proximity in position and velocity to the disk of VCC873 and the similar abundances observed 
in the two objects might indicate a galactic fountain origin for \emph{122603+130724}.
The expelled gas could have condensed into a cloud during its infall back to the disk,  
triggering star formation activity.
If galactic fountains are as frequent as claimed by Fraternali et al. (\cite{fraternali}), 
extragalactic HII regions possibly associated with this phenomenon might contribute 
significantly to the diffuse intra-cluster light.

%would be frequent in star-forming galaxies, 
%contributing significantly to the diffuse intra-cluster light.
%The galaxy fountain phenomenon is supposed to be common among 
%spiral galaxies (Fraternali et al. \cite{fraternali}). 
%Therefore, if it is connected with the formation of \emph{122603+130724},  
%extragalactic HII regions would be usually present in star-forming galaxies, 
%contributing significantly to the diffuse intra-cluster light.

%and pointing out the possibility that high metallicity extragalactic HII regions 
%could form even in non interacting systems. 
%however it has lost approximately 75 \% of its 
%original hydrogen content ($HI_{def}$=0.63, Cayatte et al. \cite{cayatte}),  
%indicating that the ram-pressure (Gunn \& Gott 1972) exerted by the dense intergalactic medium (IGM) has caused 
%its hydrogen deficiency. Thus the presence of \emph{122603+130724} 
%but also from enriched gas stripped by ram pressure.
%A possible interpretation is given by Dyson \& Hartquist (\cite{dyson})
%who argue that these systems may be formed from large-scale compressions produced by high velocity streams of gas. 
%Whatever is their origin, extragalactic HII regions could be the birth place of O-B stars observed in galactic halos 
%(Conlon et al. \cite{conlon}, Comeron et al. \cite{comeron}).

\begin{acknowledgements}      
The TACS of the Loiano and Calar Alto telescopes are acknowledged for the generous time allocation to this project. 
This work could not have been completed without the NASA/IPAC Extragalactic Database (NED) which is operated by the 
Jet Propulsion Laboratory, Caltech under contract with NASA.
We also made use of the GOLD Mine Database, operated by the Universit\`{a} degli Studi di Milano-Bicocca. 
\end{acknowledgements}

%\begin{figure*}
%\centering
% \includegraphics[width=9.5cm]{122544_130740FC.ps}
% \includegraphics[width=9.5cm]{122603_130724FC.ps}
% \includegraphics[width=9.5cm]{122620_131002FC.ps}
% \includegraphics[width=9.5cm]{123015_122812FC.ps}
% \includegraphics[width=9.5cm]{123021_121614FC.ps}
% \includegraphics[width=9.5cm]{123728_082540FC.ps}
% \caption{$r'$ band images of the six emission-line objects.}
% \label{FC}
% \end{figure*}

%\begin{figure*}
%\centering
%\includegraphics[width=19cm]{fig_FC.ps}
%\caption{$r'$ band images of the six emission-line objects.}
%\label{FC}
%\end{figure*}

\begin{table*}  
\caption{The spectrograph characteristics} 
\label{Tab1} \[ 
\begin{array}{lccccc} 
\hline 
\noalign{\smallskip}   
\rm Observatory & \rm Spectrograph & \rm Dispersion & \rm Coverage  & \rm CCD & \rm pix \\      
	&  & \rm \AA/mm & \rm \AA &  & \rm \mu m \\ 
\noalign{\smallskip} 
\hline 
\noalign{\smallskip} 
Loiano     & BFOSC & 198 & 3600-8900  & 1340\times1300~EEV & 20 \\
Calar Alto & CAFOS & 187 & 3600-10200 & 2048\times2048~SIT & 24 \\
\noalign{\smallskip} 
\hline 
\end{array} \] 
\end{table*}

\begin{table*}
\caption{Journal of the observations.}
\label{Tab2} \[ 
\begin{array}{lccc}
\hline 
\noalign{\smallskip}   
\rm Object & \rm Telescope & \rm Date & \rm Exp. Time \\
	&         &       &     (sec)  \\
\noalign{\smallskip} 
\hline 
\noalign{\smallskip}
122544+130740  & Calar~Alto & 10~Apr~02 & 1\times1800 \\
122544+130740  & Loiano     & 03~Mar~03 & 1\times1800 \\
122603+130724  & Loiano     & 12~Feb~02 & 1\times1200 \\
122603+130724  & Loiano     & 03~Feb~03	& 2\times1800 \\
122620+131002  & Loiano     & 29~Mar~03	& 3\times1200 \\
123015+122812  & Calar~Alto & 09~Apr~02	& 1\times1800 \\
123021+121614  & Calar~Alto & 09~Apr~02	& 1\times1800 \\
123728+082540  & Calar~Alto & 10~Apr~02 & 1\times1800 \\
\noalign{\smallskip} 
\hline 
\end{array} \]
\end{table*}

\begin{table*}
\caption{Spectroscopic parameters of the observed objects.}
\label{Tab3} \[ 
\begin{array}{lccccc}
\hline 
\noalign{\smallskip}   
 \rm Object & \rm R.A.  & \rm Dec.  & \rm r'  & \rm vel & \rm Emission~Lines \\      
          & \rm (J.2000) & \rm (J.2000) & \rm (mag) & \rm (km \ s^{-1}) &     \\	
\noalign{\smallskip} 
\hline 
\noalign{\smallskip}
122544+130740  & 12 25 44.20 & +13 07 40.0 & 20.10  & 80160 &  \rm [OII],[OIII] \\
122603+130724  & 12 26 02.91 & +13 07 23.7 & 21.60  &	200 &  \rm  H_{\beta}, H_{\alpha},[SII]  \\
122620+131002  & 12 26 20.80 & +13 10 02.0 & 21.19  & 80160 &  \rm [OII], [OIII] \\
123015+122812  & 12 30 15.20 & +12 28 12.0 & 21.10  & 81213 &  \rm [OII], H_{\beta}, [OIII] \\
123021+121614  & 12 30 21.10 & +12 16 14.0 & 20.70  & 81213 &  \rm [OII], [OIII], H_{\alpha} \\
123728+082540  & 12 37 28.00 & +08 25 40.0 & 20.60  & 79733 &  \rm [OII], H_{\beta}, [OIII], H_{\alpha}, [SII] \\
\noalign{\smallskip} 
\hline 
\end{array} \]
\end{table*}

\begin{table*}
\caption{Observed and reddening corrected emission-line fluxes of \emph{122603+130724}.}
\label{lines} \[ 
\begin{array}{lccc}
\hline 
\noalign{\smallskip}   
 \rm Line & \rm \lambda  & \rm ~~Observed~ Flux~~  & \rm ~~Dereddened~ Flux~~ \\      
          & \rm (\AA) & \rm (F/F_{H_{\beta}}) & \rm (F/F_{H_{\beta}})    \\	
\noalign{\smallskip} 
\hline 
\noalign{\smallskip}
\rm [OII]	 &  3727	 & < 0.23   & < 0.38   \\
\rm H_{\beta}    &  4861	 &	1   &    1   \\
\rm [OIII]	 &  4959	 & < 0.18   &  < 0.17  \\
\rm [OIII]	 &  5007	 & < 0.18   &  < 0.17  \\
\rm H_{\alpha}   &  6563	 &  4.55    &  2.86  \\
\rm [NII]	 &  6583	 &  0.78    &  0.49  \\
\rm [SII]	 & 6717+6731     &  0.65    &  0.40  \\
\noalign{\smallskip} 
\hline 
\end{array} \]
\end{table*}

\clearpage

\end{document}